# BLACK-BODY PHOTON CLUSTERING BY SEMI-CLASSICAL MEANS

J. P. LESTONE

Applied Physics Division, Los Alamos National Laboratory
Los Alamos, NM, 87545, USA

April 20$^{th}$, 2008

If stimulated emission could be turned off then only uncorrelated photons would be emitted from black bodies and the photon counting statistics would be Poissonian. Through the process of stimulated emission, some fraction of the photons emitted from a black body are correlated and thus emitted in clusters. This photon clustering can be calculated by semi-classical means. The corresponding results are in agreement with quantum theory.

*Keywords*: Black body; photon clustering; Hanbury Brown and Twiss.

It has been known for more than fifty years that the detection of two photons in closely spaced detectors is correlated in time.[1,2] These observed detector correlations can be explained by both classical[3] and quantum[4] theory. Quantum theory has been used to show that photons can be emitted from black bodies in clusters.[4,5] The probability that $j$ photons, in a narrow frequency range, are emitted in a time interval much larger than the inverse of the frequency is given by a negative binomial distribution. If this distribution is assumed, then the exchange of photons between two black-body cavities is in agreement with the entropy fluctuation theorem.[6] It is well known that stimulated emission can be deduced by considering an ensemble of excited states in thermal equilibrium with the photons emitted from a black body.[7] Stimulated emission is an essential ingredient in obtaining the correct energy distributions for photons emitted from black bodies.[8] Given that some black-body photons are produced by stimulated emission, it follows that some fraction of black-body photons are correlated, and thus emitted in clusters. In the present paper, we show that if stimulated emission is included, then semi-classical calculations give black-body photon clustering in agreement with quantum theory.

One of the simplest methods for obtaining the properties of classical particle emission from a container is to use the Bohr-Wheeler decay width[9]

$$\Gamma^{BW} = \frac{1}{2\pi} \frac{N_{TS}}{\rho(U)}, \qquad (1)$$

where $\rho(U)$ is the total level density of the system containing energy $U$, and $N_{TS}$ is the number of transition states. For simplicity, we assume a spherical container with a radius $r$ with a small hole in the surface with area $\sigma$. The number of transition states for the emission of particles with kinetic energies between $\varepsilon$ and $\varepsilon + d\varepsilon$ is

$$N_{TS} = \sum_{L=0}^{L_{MAX}} (2L+1) \frac{\sigma}{4\pi r^2} \rho(U-\varepsilon) \, d\varepsilon, \qquad (2)$$

where $L\hbar$ is the angular momentum of the emitted particle. For an emitted particle with kinetic energy $\varepsilon$ and mass $m$, the maximum angular momentum squared is given by

$$L_{MAX}^2 \hbar^2 = 2m\varepsilon r^2, \qquad (3)$$

Assuming a large classical system where $L_{MAX} \gg 1$, and using the definition of the temperature as the inverse derivative of the natural log of the system's level density, gives the probability per unit time that a particle is emitted from a container

$$\lambda(\varepsilon) = \frac{\Gamma^{BW}(\varepsilon)}{\hbar} = \frac{2m\varepsilon\sigma}{8\pi^2 \hbar^3} e^{-\varepsilon/T} d\varepsilon. \qquad (4)$$

The temperature is expressed in units of energy. Using the well-known result for the number density of ideal particles in a box

$$n = (mT/2\pi\hbar^2)^{3/2} \qquad (5)$$

gives

$$\lambda(\varepsilon) = \frac{\sigma n \varepsilon \, d\varepsilon}{T\sqrt{2\pi mT}} e^{-\varepsilon/T}. \qquad (6)$$

Using Eq. (6) and assuming that the particles move independently of each other, it has been shown that the equilibration of two ideal gases via a small hole in a separating wall is in agreement with the entropy fluctuation theorem.[6]

The classical emission of photons from a black body can be obtained by replacing the square of the momentum of a classical particle $2m\varepsilon$ in Eq. (4) with the corresponding value for photons $(\varepsilon/c)^2$. Multiplying by two to include both states of helicity gives

$$\lambda_c(\omega) = \frac{\sigma \omega^2 \, d\omega}{(2\pi c)^2} e^{-\varepsilon/T}. \qquad (7)$$

The same result can be obtained using the method to calculate particle evaporation developed by Weisskopf.[10,11] Using the results of ref 6, and assuming only the classical emission of uncorrelated photons via Eq. (7), it is simple to show that the equilibration of two classical black-body cavities is in agreement with the entropy fluctuation theorem. However, the result in Eq. (7) differs from the full result for black-body radiation because it only represents the classically emitted (spontaneous) photons and does not include stimulated emission. For a system in thermodynamic equilibrium, the total emission of photons relative to the spontaneously emitted photons is

$$\frac{\text{total}}{\text{spontaneous}} = \frac{e^{\varepsilon/T}}{e^{\varepsilon/T}-1}. \qquad (8)$$

Applying this correction factor to the classical emission given by Eq. (7) gives the full black-body result

$$\lambda_b(\omega) = \frac{\sigma \omega^2 \, d\omega}{(2\pi c)^2} \frac{1}{e^{\varepsilon/T} - 1}. \tag{9}$$

Using Eq. (9) and assuming the photons to be independent of each other gives results in disagreement with the entropy fluctuation theorem.[6] Agreement with the fluctuation theorem can be obtained if photon clustering is taken into account.[6] Using quantum theory, it can be shown that the probability that $j$ photons in a narrow frequency range are emitted in a time interval $t$ much larger than the inverse of the frequency is given by a negative binomial distribution[5,6]

$$P(j) = \frac{\Gamma(j + \lambda_c t/f)}{j! \, \Gamma(\lambda_c t/f)} (1-f)^{\lambda_c t/f} f^j, \tag{10}$$

where $\lambda_c t$ is the expected number of photons if all stimulated emission could be turned off and $f = e^{-\varepsilon/T}$ is the stimulated emission relative to the photon absorption in the surface of the black body. The negative binomial distribution is necessary for the exchange of photons between two black-body cavities to be consistent with the entropy fluctuation theorem.[6] It is useful to define the photon multiplicities distribution

$$M(j) = \lim_{\lambda_c t \to 0} \frac{P(j)}{\lambda_c t} = \frac{f^{(j-1)}}{j}. \tag{11}$$

In the limit $f \to 0$, $M(1)=1$ and $M(j)=0$ for all $j > 1$. Therefore, in the limit $f \to 0$ there are no photon-photon correlations and the counting statistics become Poissonian. The multiplicity distribution is

$$M(j), \, j = 1, 2, 3, 4 \ldots = 1, \, f/2, \, f^2/3, \, f^3/4 \ldots \tag{12}$$

The average number of photons per emission event is therefore given by

$$\bar{j} = 1 + f + f^2 + \ldots = \frac{1}{1-f} = \frac{e^{\varepsilon/T}}{e^{\varepsilon/T} - 1}. \tag{13}$$

This is the same as the ratio given by Eq. (8).

If stimulated emission is included semi-classically, then photon-photon correlations can be obtained without invoking full quantum theory. Even though photon clustering can be derived easily from quantum theory,[4] it is of interest to see if the semi-classical correlations are in agreement with full quantum theory. The properties of black-body radiation are often determined by invoking a surface that absorbs all radiation that falls onto it, without any reference to the microscopic processes responsible for the complete absorption. Here, we assume a non-scattering material that is much thicker than the mean free path for absorption. The non-scattering assumption is essential in obtaining a perfect black body. If a finite scattering cross section were allowed, then some fraction of the incident radiation could make it back to the surface. Consider a spontaneously emitted photon generated at a point inside a black body with a path length $x_o$ to the surface. It is convenient to work in units of length equal to the photon's mean free path assuming only absorption (no stimulated emission). Excluding the possibility of stimulated emission, the photon is either emitted with a probability $\exp(-x_o)$ or absorbed with a probability $1-\exp(-x_o)$, and no photon-photon correlations are generated. If stimulated emission is included, then there exists a probability that there are $i$ interaction locations along the path to the surface, and that some fraction of the seed spontaneously emitted photons lead to multiple correlated photons at the black body's surface.

It is easy to simulate the passage of photons through the surface of a black body by Monte-Carlo means. We have done this assuming the mean free path is 1 if stimulated emission is excluded as a possible interaction. Including stimulated emission, the mean free path is $1/(1+f)$. At each interaction site, each photon has an absorption probability $q=1/(1+f)$ and a stimulated emission probability $p=f/(1+f)$. Each Monte-Carlo event is started with a single spontaneously emitted photon with a path length from the surface uniformly chosen from 0 to 20. These seed events are chosen randomly in time with a probability per unit time of $20\lambda_c$. Figures 1 and 2 display numerical results obtained by simulating $10^7$ time intervals of $0.01\lambda_c$ and $5\lambda_c$, respectively, both with $f=0.7$. The solid lines show the corresponding quantum theory results obtained using Eq. (10). The excellent agreement between the semi-classical Monte-Carlo results and quantum theory suggests the agreement is exact and that an analytical solution for the semi-classical model exists.

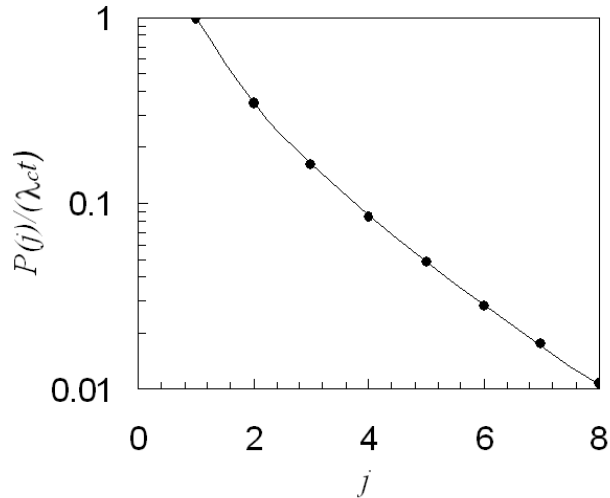

Fig. 1. $P(j)/(\lambda_c t)$ with $f=0.7$ and $t=0.01\lambda_c$. The symbols show the results obtained by the Monte-Carlo simulation of $10^7$ time intervals. The solid curve is the corresponding quantum result obtained via Eq. (10).

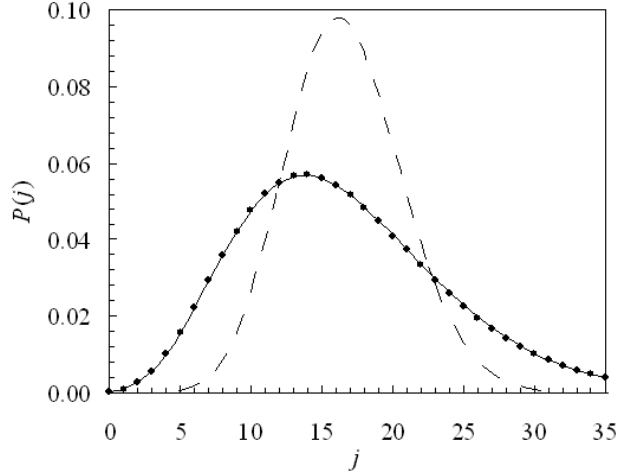

Fig. 2. $P(j)$ with $f=0.7$ and $t=5\lambda_c$. The symbols show the results obtained by the Monte-Carlo simulation of $10^7$ time intervals. The solid curve is the corresponding quantum result obtained via Eq. (10). The dashed curve shows a Poisson distribution with the same mean.

To obtain an analytical solution to the semi-classical transport of photons through the surface of a black body, we first consider a spontaneously emitted photon originating a path length $x_o$ from the surface. We consider the probability that $i$ interactions occur at distances $x_1, x_2, \ldots x_i$ from the surface, and that the number of photons after each interaction is $n_1, n_2, \ldots n_i$. This probability can be written as

$$Q(x_0, x_1, \ldots x_i, n_1, \ldots n_i) = dx_1\, dx_2 \ldots dx_i \times \prod_{m=1}^{i} e^{-(x_{m-1}-x_m)n_{m-1}(1+f)}\, n_{m-1}\, s_m\, e^{-x_i n_i (1+f)} \quad (14)$$

where $n_0=1$ and the $s_m$ are given by $s_m=f$, if $\Delta n_m = n_m - n_{m-1} = +1$ and $s_m=1$, if $\Delta n_m = -1$. $\Delta n_m = +1$ and $-1$ correspond to stimulated emission and photon absorption, respectively. To obtain the probability per unit time that $j=n_i$ photons are emitted after $i$ interactions with $n_m$ photons following each interaction, we multiply Eq. (14) by $\lambda_c$ and integrate over all $x_m$ ($m=0$ to $i$),

$$Q(n_1, \ldots n_i) = \int_{x_0=0}^{\infty} \int_{x_1=0}^{x_0} \ldots \int_{x_i=0}^{x_{i-1}} \lambda_c\, Q(x_0, x_1, \ldots x_i, n_1, \ldots n_i)\, dx_0$$

$$= \prod_{m=1}^{i} s_m \prod_{m=1}^{i-1} n_m \frac{\lambda_c}{(1+f)^{i+1}} \frac{1}{\prod_{m=1}^{i} n_m} = \frac{\lambda_c q^{i+1}}{n_i}\left(\frac{p}{q}\right)^a \quad (15)$$

where $a=(j+i-1)/2$ is the total number of stimulated emissions along the exit path. The probability per unit time that $j$ correlated photons are emitted can be obtained by multiplying Eq. (15) by a combinatorial factor associated with all possible $n_1$, $n_2$,..., $n_{i-1}$, $n_i=j$ for a given number of interactions $i$ and then by summing over all allowable $i$. The corresponding result is

$$Q(j) = \frac{\lambda_c q}{j} \sum_{i=j-1}^{\infty} [i,j] \, q^{(i-j+1)/2} \, p^{(j+i-1)/2}, \tag{16}$$

where $[i,j]$ is the number of ways of arranging $n_1$, $n_2$, ... $n_{i-1}$, $n_i=j$, given $i$ interaction sites.

The passage of photons through a black body is closely related to an asymmetric random walk starting at $j=1$ with probabilities of a unit step to higher and lower $j$ equal to $p$ and $q$, respectively, and with an absorptive boundary at $j=0$. If the system gets to $j=0$, the walk is terminated and no additional steps are taken. The probabilities of getting to $j$ after $i<10$ steps are listed in Table I. The combinatorial factors in front of the $q^{(i-j+1)/2} p^{(j+i-1)/2}$ terms listed in Table 1 are the same combinatorial factors needed to evaluate Eq. (16), and we write the random walk probabilities as

$$W(i,j) = [i,j] \, q^{(i-j+1)/2} \, p^{(j+i-1)/2}. \tag{17}$$

Table I. The probability of getting to $j$ after $i<10$ steps in an asymmetric random walk starting at $j=1$ with an absorptive boundary at $j=0$.

| $j$ | Probability $W(i,j)$ | | | | | | | | |
|---|---|---|---|---|---|---|---|---|---|
| 10 | | | | | | | | | $p^9$ |
| 9  | | | | | | | | $p^8$ | |
| 8  | | | | | | | $p^7$ | | $8qp^8$ |
| 7  | | | | | | $p^6$ | | $7qp^7$ | |
| 6  | | | | | $p^5$ | | $6qp^6$ | | $27q^2p^7$ |
| 5  | | | | $p^4$ | | $5qp^5$ | | $20q^2p^6$ | |
| 4  | | | $p^3$ | | $4qp^4$ | | $14q^2p^5$ | | $48q^3p^6$ |
| 3  | | $p^2$ | | $3qp^3$ | | $9q^2p^4$ | | $28q^3p^5$ | |
| 2  | $p$ | | $2qp^2$ | | $5q^2p^3$ | | $14q^3p^4$ | | $42q^4p^5$ |
| 1  | 1 | $qp$ | | $2q^2p^2$ | | $5q^3p^3$ | | $14q^4p^4$ | |
| 0  | | $q$ | $q^2p$ | | $2q^3p^2$ | | $5q^4p^3$ | | $14q^5p^4$ |
| $i$ | 0 | 1 | 2 | 3 | 4 | 5 | 6 | 7 | 8 | 9 |

For a system in thermo-dynamical equilibrium, the probability that a given photon will induce a stimulated emission is less than the probability that it will be absorbed and thus $p$ is less than $q$. Applying this constraint to the asymmetric random walk leads to the result that in the limit as the number of steps $i \to \infty$ all paths end at $j=0$. We can therefore write

$$\sum_{i=1}^{\infty} W(i,0) = \sum_{i=1}^{\infty} [i,0] \, q^{(i+1)/2} \, p^{(i-1)/2} = 1. \tag{18}$$

Properties of the combinatorial factors $[i,j]$ include

$$[0,1] = 1. \tag{19}$$

$$[i,j] = 0 \quad \text{if } i < j-1 \text{ or}$$
$$\quad \text{if both } i \text{ and } j \text{ are even or} \tag{20}$$
$$\quad \text{if both } i \text{ and } j \text{ are odd.}$$

$$[j-1, j] = 1. \tag{21}$$
$$[i+1, j] = [i, j-1] + [i, j+1] \quad \text{if } j \geq 2. \tag{22}$$
$$[i, 2] = [i+1, 1] = [i+2, 0] \quad \text{if } i \geq 1. \tag{23}$$
$$[i, 1] = [i+1, 0] \quad \text{if } i \geq 0. \tag{24}$$

From Eq. (16) we can write the probability per unit time that a single uncorrelated photon is emitted from a black body as

$$Q(1) = \lambda_c q \sum_{i=0}^{\infty} [i, 1] \, q^{i/2} p^{i/2} \, . \tag{25}$$

By substituting [$i$,1] for [$i$+1,0] and changing the summation variable by 1 we get

$$Q(1) = \lambda_c \sum_{i=1}^{\infty} [i, 0] \, q^{(i+1)/2} p^{(i-1)/2} = \lambda_c \, . \tag{26}$$

For Eq. (18) we see that the summation in Eq. (26) is equal to 1 and therefore the probability for the emission of uncorrelated photons is the same as the purely classical result. From Eq. (16) we write the probability per unit time that a pair of correlated photons is emitted from a black body as

$$Q(2) = \frac{\lambda_c q}{2} \sum_{i=1}^{\infty} [i, 2] \, q^{(i-1)/2} p^{(i+1)/2} \, . \tag{27}$$

By substituting [$i$,2] for [$i$+2,0] and changing the summation variable by 2 we get

$$Q(2) = \frac{\lambda_c}{2q} \sum_{i=3}^{\infty} [i, 0] \, q^{(i+1)/2} p^{(i-1)/2}$$
$$= \frac{\lambda_c}{2q} \left\{ \sum_{i=1}^{\infty} [i, 0] \, q^{(i+1)/2} p^{(i-1)/2} - q \right\} = \frac{\lambda_c f}{2} \, . \tag{28}$$

To obtain an analytical expression for $Q(j>2)$, a more complex approach is needed. We define

$$Z(n) = \frac{nQ(n)}{\lambda_c} = q \sum_{i=n-1}^{\infty} [i, n] \, q^{(i-n+1)/2} p^{(n+i-1)/2} \, . \tag{29}$$

Using Eqs (21) and (22), it can be shown that

$$Z(n) = q \, Z(n+1) + p \, Z(n-1) \, , \text{ and thus} \tag{30}$$

$$Z(n+1) = \frac{Z(n)}{q} - \frac{p}{q} Z(n-1) \, . \tag{31}$$

From Eqs (26) and (28) we can write

$$Z(n = 1, 2) = f^{n-1} \, . \tag{32}$$

If we substitute $Z(n) = f^{n-1}$ and $Z(n-1) = f^{n-2}$ into Eq. (31) we obtain

$$Z(n+1) = f^n \, , \tag{33}$$

and therefore, by induction, the probability per unit time that $j$ correlated photons are emitted from a black body is

$$Q(j) = \frac{\lambda_c f^{j-1}}{j} \, . \tag{34}$$

The mean number of photons emitted in a time interval $t$ is therefore given by

$$\bar{j} = \lambda_c t \left(1 + f + f^2 + ...\right) = \frac{\lambda_c t \, e^{\varepsilon/T}}{e^{\varepsilon/T} - 1}, \qquad (35)$$

in agreement with the quantum result. In the limit as $\lambda_c t \to 0$, no two emission events will contribute to the same time interval, and the semi-classical multiplicity distribution is given by

$$M(j) = \frac{Q(j)}{\lambda_c} = \frac{f^{j-1}}{j}, \qquad (36)$$

in agreement with quantum theory. Given that the semi-classical results for the mean number of photons emitted in a given time interval $t$ and the multiplicity distribution are in agreement with quantum theory, it follows that the semi-classical counting statistics will be as given by Eq. (10).

The photon clustering from black bodies, discussed here, will lead to both photon-photon temporal and spatial correlations. The spatial correlations could be weakened for partial black bodies via photon scattering in the emitter's surface. The spatial correlations will depend on the angular size of the black body in a manner analogous to the Hanbury Brown and Twiss effect. To calculate the cross correlation of the signals between two detectors observing photons at the same frequency, we assume detectors 1 and 2 are both a large distance $d$ from a black-body source each covering a solid angle $d\Omega = d\Omega_1 = d\Omega_2$, and that the separation between these two detectors $\Delta r$ is much smaller than the distance $d$ and thus the cross sectional area of the source as viewed by both detectors is the same value $\sigma$. The mean rate of photon detection in each of the detectors is given by

$$\bar{n}_i = \lambda_b \frac{d\Omega_i}{\pi}. \qquad (37)$$

If we assume that on the detection of a single photon, a unit current is produced for a time duration $dt$, then the average detector signal will be

$$\bar{S}_i = \lambda_b \frac{d\Omega_i}{\pi} dt. \qquad (38)$$

The cross correlation between two detectors can be divided into two parts. One part is associated with the accidental coincidences between uncorrelated photons and the other is associated with true coincidences between correlated photons. The cross correlation associated with accidental coincidences is given by the average rate of detections in detector 1, multiplied by the probability of a random overlapping detection in detector 2, multiplied by the average signal associated with a single accidental coincidence, and can be expressed as

$$<S_1 S_2>_a = \lambda_b \frac{d\Omega_1}{\pi} \times \lambda_b \frac{d\Omega_2}{\pi} 2 dt \times \frac{dt}{2} = \left(\frac{\lambda_b \, d\Omega \, dt}{\pi}\right)^2 = \bar{S}_1 \bar{S}_2. \qquad (39)$$

The cross correlation due to correlated photons is more complex. We shall, at first, determine the cross correlation signal generated by emission events involving a photon multiplicity of $m$. The probability per unit time that exactly $m$ correlated photons are emitted from a black body is governed by Eq. (34). The probability that any of the $m$ photons encounters either of the two detectors is given by $P_1 = 2m \cdot d\Omega/\pi$. If we assume that the $m$ correlated photons travel independently of each other, then given that one of the detectors has observed one of the $m$ correlated photons, the probability that one of the other $m-1$ photons is seen in the other detector is governed by diffraction and given by

$$P_2 = (m-1)\frac{\sigma}{\lambda^2} F(r_1, r_2) d\Omega, \qquad (40)$$

where $r_1$ and $r_2$ are vectors pointing from the source to the two detectors. The function $F$ is the diffraction pattern associated with an aperture equal to the cross section of the source as viewed from the detectors. The properties of $F$ include[12] $F(r_1, r_1) = 1$; $F(r_1, r_2) = 0$ if the two detectors are separated by a distance much larger than the wavelength divided by the angular size of the source; and

$$\iint \frac{\sigma}{\lambda^2} F(r_1, r_2) d\Omega = 1. \qquad (41)$$

Given the above discussion, and assuming $dt$ is much larger than the coherence time $\tau = 1/d\nu = 2\pi/d\omega$, then the cross correlation signal due to the emission of events involving a photon multiplicity of $m$ is given by

$$<S_1 S_2>_m = \frac{\lambda_c f^{m-1}}{m} 2m \frac{d\Omega}{\pi} (m-1) \frac{\sigma}{\lambda^2} F(r_1, r_2) d\Omega\, dt. \qquad (42)$$

Summing over all $m>1$ gives the cross correlation signal due to all correlated photons

$$<S_1 S_2>_c = 2\lambda_c f \frac{d\Omega}{\pi} \frac{\sigma}{\lambda^2} F(r_1, r_2) d\Omega\, dt \sum_{m=2}^{\infty} (m-1) f^{m-2}$$

$$= 2\lambda_c f \frac{d\Omega}{\pi} \frac{\sigma}{\lambda^2} F(r_1, r_2) d\Omega\, dt \left( \sum_{m=0}^{\infty} f^m \right)^2$$

$$= \left( \frac{\lambda_b d\Omega\, dt}{\pi} \right)^2 \frac{2\pi f}{\lambda_c dt} \frac{\sigma}{\lambda^2} F(r_1, r_2) = \bar{S}_1 \bar{S}_2 \frac{\tau}{dt} F(r_1, r_2). \qquad (43)$$

Adding the signals from both accidental and correlated sources gives the total cross correlation

$$<S_1 S_2> = \bar{S}_1 \bar{S}_2 \left(1 + \frac{\tau}{dt} F(r_1, r_2)\right). \qquad (44)$$

This is the same cross correlation as for the Hanbury Brown and Twiss effect.[13]

The semi-classical calculation described here assumes photons act independently of each other and that the clustering of photons is determined by stimulated emission. It is interesting that this approach can reproduce the quantum theory result for the clustering of photons from black bodies and that this clustering gives the same cross correlation signal as the Hanbury Brown and Twiss effect.